# Recursive Subdivision of Urban Space and Zipf's law

Yanguang Chen[1], Jiejing Wang[2]

(1. Department of Geography, College of Urban and Environmental Sciences, Peking University, Beijing 100871, PRC; 2. Department of Urban Planning and Design, The University of Hong Kong, Pokfulam Road, Hong Kong SAR)

**Abstract**: Zipf's law can be used to describe the rank-size distribution of cities in a region. It was seldom employed to research urban internal structure. In this paper, we demonstrate that the space-filling process within a city follows Zipf's law and can be characterized with the rank-size rule. A model of spatial disaggregation of urban space is presented to depict the spatial regularity of urban growth. By recursive subdivision of space, an urban region can be geometrically divided into two parts, four parts, eight parts, and so on, and form a hierarchy with cascade structure. If we rank these parts by size, the portions will conform to the Zipf distribution. By means of GIS technique and remote sensing data, the model of recursive subdivision of urban space is applied to three cities of China. The results show that the intra-urban hierarchy complies with Zipf's law, and the values of the rank-size scaling exponent are very close to 1. The significance of this study lies in three aspects. First, it shows that the strict subdivision of space is an efficient approach to revealing spatial order of urban form. Second, it discloses the relationships between urban space-filling process and the rank-size rule. Third, it suggests a new way of understanding fractals, Zipf's law, and spatial organization of urban evolution.

## 1. Introduction

An urban system can be divided into two levels: one is cities as systems, and the other, systems of cities (Berry, 1964). Where academic fields are concerned, the former belongs to the intra-urban geography (micro level), and the latter, the interurban geography (macro level). A city as a system is defined for individual cities and, sometimes, called 'city system' (Batty and Longley, 1994), while a system of cities is often defined as a network of cities and its hinterland in a region (Mayhew, 1997). In a broad sense, all cities which are connected by transport and communication networks within a geographical region compose a system of cities (Bourne and Simons, 1978; Chen, 2008). If the largest city in a region (e.g., a country) is a megacity or a world city with a scope of influence going beyond the region to a great extent, the cities often accord with what is called *primate distribution*. Otherwise, the cities comply with the *rank-size distribution* and follow Zipf's law (Zipf, 1949). In urban geography, Zipf's law is always used to describe the size distribution of cities in an area, and it is often associated with the concept of systems of cities.

The Zipf distribution is one of the ubiquitous empirical observations across the individual sciences, which cannot be understood within the set of references developed within the specific scientific domain (Bak, 1996). Zipf's law originated from urban studies (Carroll, 1982), but it is frequently observed in the natural living world as well as in social institutions (Altmann *et al*, 2009; Axtell, 2001; Bettencourt *et al*, 2007; Blasius *et al*, 2009; Cancho and Solé, 2003; Carlson and Doyle, 1999; Ferrer i Cancho *et al*, 2005; Furusawa and Kaneko, 2003; Gabaix, 2009; Newman, 2005; Petersen *et al*, 2010; Podobnik *et al*, 2010; Shao *et al*, 2011; Stanley *et al*, 1995). For urban studies, Zipf's law is always employed to describe the scaling relation between rank and size of cities in a region (Batty, 2006; Carroll, 1982; Chen, 2008; Córdoba, 2008; Gabaix, 1999; Gabaix and Ioannides, 2004; Gangopadhyay and Basu, 2009; Jiang and Yao, 2010; Manrubia and Zanette, 1998; Moura and Ribeiro, 2006; Peng, 2010; Zanette and Manrubia, 1997). The rank-size patterns of street hierarchies show that the length of streets conforms to Zipf's law (Jiang, 2009). However, so far, the rank-size law has been principally applied to elements rather than structure of a system. On the other hand, it is imperative to develop theories and methods for spatial analysis within a city in order to improve city planning and optimize urban structure. Fractal geometry and the correlated percolation model (CPM) have been used to explore urban structure and form (Batty, 2008; Batty and Longley, 1994; Benguigui *et al*, 2000; Feng and Chen, 2010; Frankhauser, 1994; Makse *et al*, 1995; Makse *et al*, 1998; Rozenfeld *et al*, 2011; Shen, 2002; Thomas *et al*, 2008; Thomas *et al*, 2010; White and Engelen, 1994). From these studies, a speculation can be aroused that the space-filling process of a city's growth is related to the rank-size distribution and conforms to Zipf's law.

Among various scientific concepts, three ones are important and can be employed to link intra-urban structure and interurban network, that is, fractal, hierarchy, and correlation. Zipf's law suggests fractal distributions and hierarchical structure (Frankhauser, 1994; Mandelbrot, 1983). A fractal is in fact a hierarchy and fractal scaling relations can be regarded as generalized correlation functions (Vicsek, 1989). The grounds that urban structure may follow Zipf's law are as below. First, if cities within a region satisfy the rank-size distribution, it can be organized into a hierarchy with cascade structure (Chen, 2012a). On the other hand, urban structure can be disaggregated into hierarchies (Batty and Longley, 1994). Second, both intra-urban structure and interurban network can be described with fractal geometry (Chen, 2008). In a sense, a city can be treated as the reduction of a system of cities. Third, the fractal models of urban form are spatial correlational function, while Zipf's law indicates a hierarchical correlational function (Chen, 2011). To connect intra-urban structure with fractals and hierarchies, we need a technique of spatial analysis, i.e., the *recursive subdivision of space* (Batty and Longley, 1994; Goodchild and Mark, 1987). By strict subdivision, urban space can be disaggregated into an urban hierarchy, and thus Zipf's law may be applied to analyzing urban space-filling process. The problem is how to testify the speculation by means of observational data of real cities.

This work is devoted to revealing the rank-size patterns of intra-urban structure using the ideas from recursive subdivision of geographical space. The study result is helpful for us to understand the Zipf distribution as well as urban evolution in the right perspective. The rest parts of the paper are organized as follows. In Section 2, a model of spatial disaggregation of urbanized area is presented. The model is based on the concepts of fractals, hierarchy, and network structure. In Section 3, three cities of China, Shanghai, Nanjing, and Hangzhou are taken as examples to make empirical analysis. The datasets from the three cities in three years will support the speculation that the intra-urban structure follows Zipf's law. In Section 4, several related questions are discussed. Finally, the paper is concluded with a brief summary of this study.

## 2. Theoretical model

The recursive subdivision of space is one of the processes of spatial disaggregation, and it can be associated with the ideas from fractals and fractal dimension (Goodchild and Mark, 1987). In theory, the strict spatial subdivision results in self-similar network, which is mathematically equivalent to the hierarchy with cascade structure (Batty and Longley, 1994; Chen, 2012b). In fact, spatial disaggregation comprises three processes: strict subdivision, hierarchy, and network structure (Figure 1). In the real world, the regular network structure with self-similarity is always changed by chance factors, but scaling invariance of the network can be found through Zipf's law (Chen, 2012c). In principle, the strict subdivision method can be applied to both the intra-urban space of an individual city and the interurban space of a system of cities. The central place models of Christaller (1933/1966) are just the self-similar networks with the recursive subdivision of geographical space (Batty and Longley, 1994).

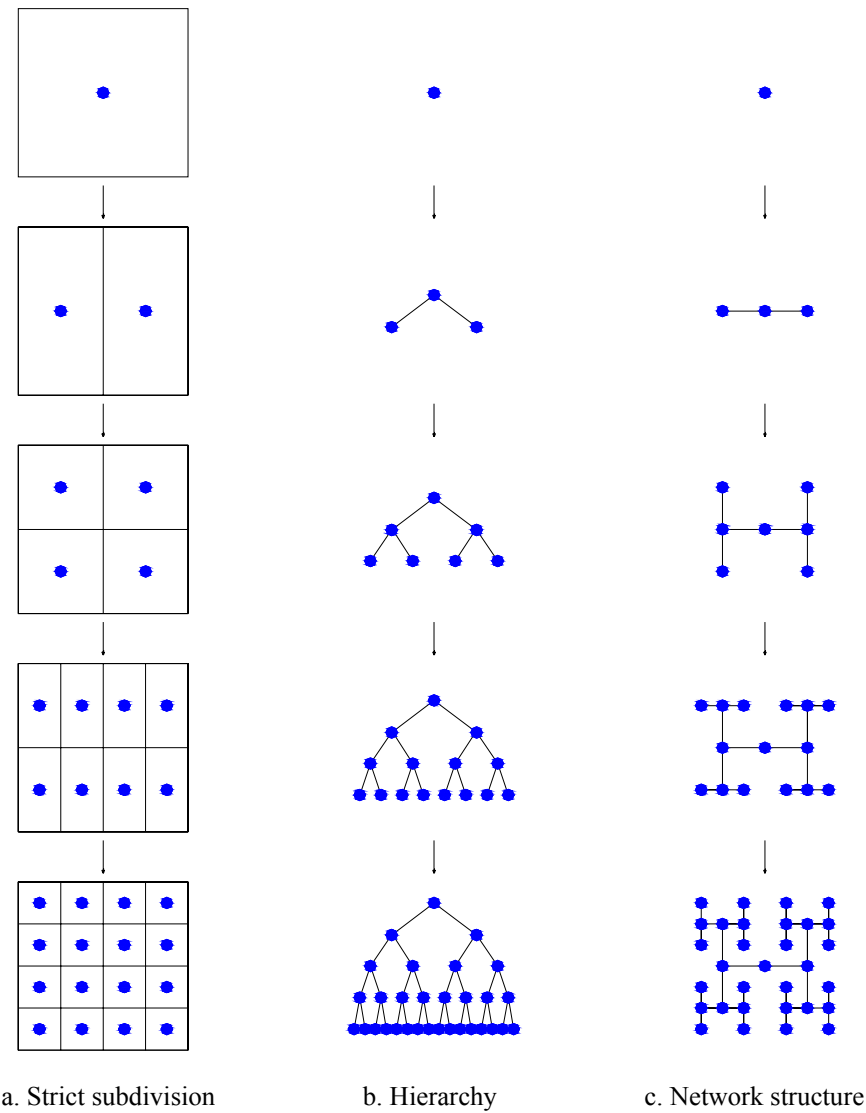

      a. Strict subdivision      b. Hierarchy      c. Network structure

**Figure 1 Spatial disaggregation: strict subdivision, hierarchy and network structure (by Batty and Longley, 1994, Page45)**

The concept of spatial disaggregation has been theoretically applied to the internal space of individual cities by Batty and Longley (1994). However, the mathematical relations have not yet been clarified, and the empirical analysis has not yet been made for urban hierarchies. The internal spatial structure in a city is expected to correspond to the spatial structure of a system of cities in which the city exists. If so, the result of spatial disaggregation should follow Zipf's law and take on hierarchical scaling. The recursive subdivision of space of a city can be illustrated with a simple example. In urban geography, a closed urban boundary is termed *urban envelope* (Batty and Longley, 1994; Longley *et al*, 1991). Suppose there is urban envelope within which the area is about 1515.5368 km$^2$ (Figure 2). The procedure of spatial subdivision is as follows. **Step 1: draw a rectangle which contains the urban envelope closely.** The rectangular frame is 55.4326 kilometers long from east to west, and 52.54 kilometers wide in south-north direction (Figure 2a). **Step 2: divide the rectangle into two equal regions using a vertical line.** Thus the urban area falls into two parts. The area of the larger one is about 858.8505 km$^2$, and that of the smaller one, 656.6863 km$^2$ (Figure 2b). **Step 3: add a horizontal line to the figure and divide the rectangle into four equal regions.** Then the urban area divides into four parts, and the areal values of these parts are 450.4168 km$^2$, 408.4337 km$^2$, 334.0249 km$^2$, and 322.6614 km$^2$, respectively. It should be noted that the lower right region actually consists of two separate pieces (Figure 2c). The area is counted in terms of each rectangular region. **Step 4: add two vertical lines to the figure and divide the rectangle into eight equal regions.** The areal values are 329.6756 km$^2$, 301.4626 km$^2$,…, and 102.3028 km$^2$ (Figure 2d). The rest steps may be implemented by analogy. In each step, the rectangle is divided into $2^n$ equal regions, and urban space fall into $2^n$ different parts ($n$=0, 1, 2, 3, …). The results can be organized as a hierarchy with cascade structure (Table 1).

Table 1 The results of strict spatial subdivision of the urban area in the first four steps

| Step | Area (sq.km.) | Average area (sq.km.) |
| --- | --- | --- |
| 1 | 1515.5368 | 1515.5368 |
| 2 | 858.8505, 656.6863 | 858.8505 |
| 3 | 450.4168, 408.4337, 334.0249, 322.6614 | 450.4168 |
| 4 | 329.6756, 301.4626, 220.3586, 208.2181, 125.8068, 120.7412, 106.9710, 102.3028 | 329.6756 |

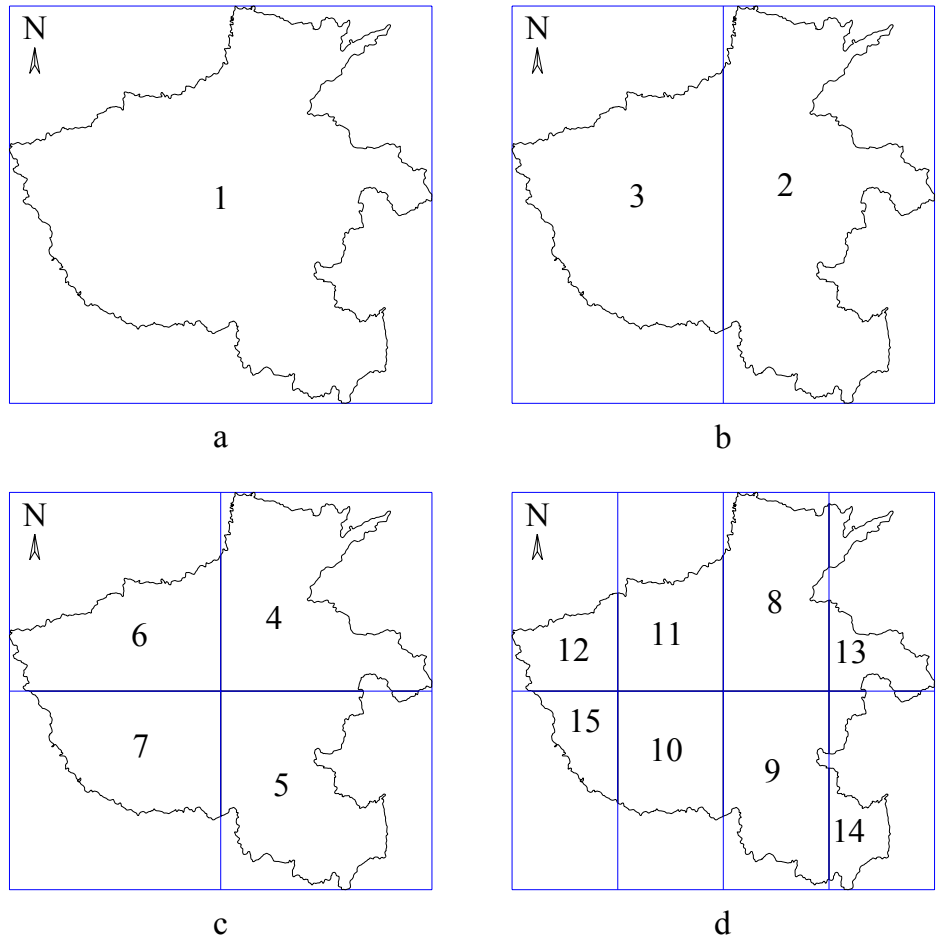

**Figure 2 A sketch map of the recursive subdivision of geographical space with an urban envelope**

**Note**: Only the first four steps are shown here. Upward direction represents north, and downward direction represents south.

The Zipf distribution is a signature of hierarchical scaling relations (Chen, 2012c). That is, if a hierarchy is self-similar and follows the scaling law on cascade structure, it will present a rank-size distribution. For a system of cities, Zipf's law states that, if the population size of a city is multiplied by its rank to the power of $q$, the product will equal the population size of the highest ranked city. Generally speaking, the power exponent $q$ is close to 1. Figure 2 exhibits a process of the recursive subdivision of urban space. According to the mathematical principle illustrated by Figure 1, this spatial disaggregation will result in a self-similar hierarchy, which is expected to comply with Zipf's law. The strict spatial subdivision gives many parts: one part in the first step, two parts in the second step, four parts in the third step, and eight parts in the fourth step, and so on. Now, conceal the hierarchical frame, and put all these parts together and rank them by areal size (Figure 3). If these parts follow Zipf's law, the area of a part ranked $k$ will be $1/k^q$th of the area of the whole city. The mathematical form of Zipf's law is

$$A_k = A_1 k^{-q}, \tag{1}$$

where $A_k$ refers to the area of the part ranked $k$, $A_1$ to the area of the whole, and $q$ to the scaling exponent. For the first four steps of the spatial subdivision of urban area shown in Figure 2, we have 15 parts. Rearranging the data displayed in Table 1 and fitting them to equation (1) yields $A_k=1815.9259k^{-0.972}$, the goodness of fit is about $R^2=0.9338$.

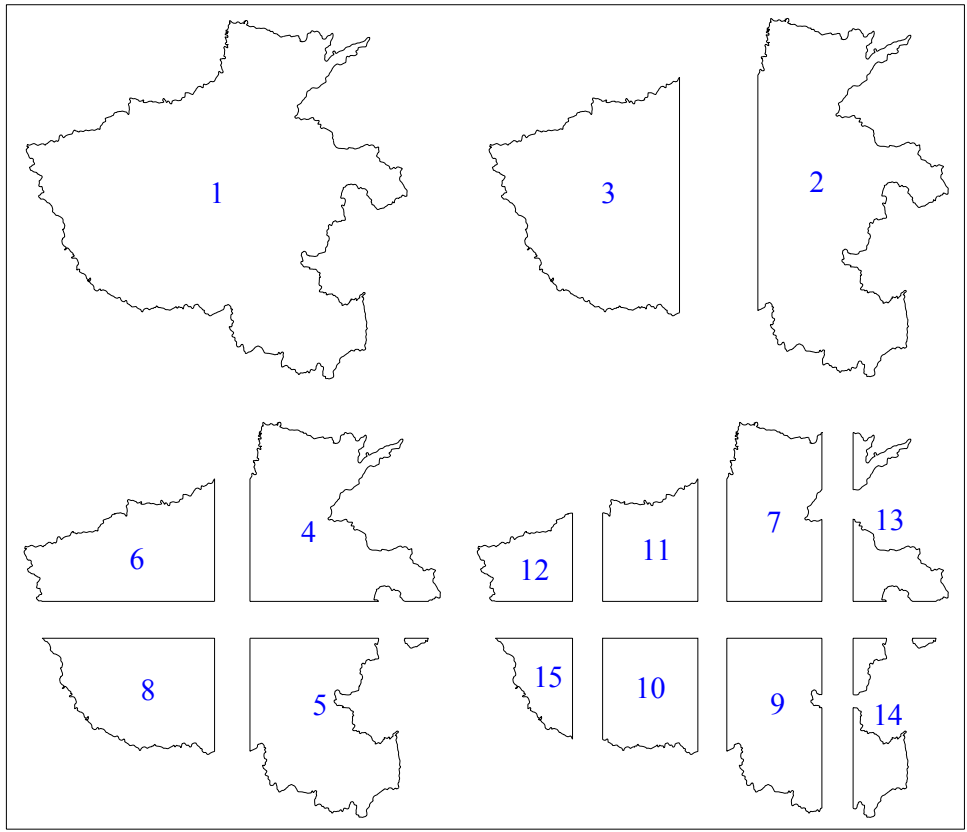

**Figure 3 The result of ranked parts from the recursive subdivision of urban space**

In theory, if the size distribution of elements in a system follows Zipf's law, the elements form or can be organized into a self-similar hierarchy (Chen, 2012a). The hierarchy of elements consisting of $M$ classes (levels) in a top-down order can be described with the discrete expressions of two exponential functions:

$$f_m = f_1 r_f^{m-1}, \tag{2}$$

$$A_m = A_1 r_a^{1-m}, \tag{3}$$

where $m=1, 2, \ldots, M$ refers to the order of classes in the hierarchy ($M$ is a positive integer), $f_m$ and $A_m$ denote the element number and average size of elements in the $m$th class, $f_1$ and $A_1$ are the number and mean size of elements in the top class, $r_f=f_{m+1}/f_m$ and $r_a=A_m/A_{m+1}$ are the *number ratio* and *size ratio* of elements, respectively. If $r_f=2$ as given, then the value of $r_a$ can be calculated; if $r_a=2$ as given, the value of $r_f$ can be derived (Davis, 1978). From Equations (2) and (3) follows the hierarchical scaling relation such as

$$f_m = \mu A_m^{-D}, \tag{4}$$

in which $\mu=f_1 A_1^D$ refers to proportionality coefficient, and $D=\ln r_f/\ln r_a$ to the fractal dimension of the self-similar hierarchies and the rank-size distributions (Chen, 2012b). For the first four levels of the hierarchy shown in Figure 2, the number law is $f_m=2^{m-1}$, where $f_1=1$ and $r_f=2$ are *ad hoc* given. Accordingly, the size law can be obtained by the least squares computation, which yields $A_m=2446.1432\exp(-0.5222m)\approx1451.1345*1.6857^{1-m}$. The goodness of fit is around $R^2=0.9821$, and the fractal parameter is estimated as $D=\ln(2)/\ln(1.6857)=1.3274$.

The simple example shown in Figure 2 and Table 1 is just used to illustrate the method of strict spatial subdivision. The defects of this instance are as below: First, this is an imaginary city from the possible world rather than a genuine city from the real world. Second, the sampling size is too small, only four levels and fifteen data points are taken into account. Third, the discontinuous and intermittent parts such as open space within the urban space are not considered. Please note that the real urban form is more complex and irregular than the model of urban area displayed in Figure 2. Actually, an urban agglomeration is not always encompassed with an envelope due to exclaves. Therefore, the case is only for teaching and cannot be taken as evidence to support the models of intra-urban hierarchical scaling. In next section, we will research the internal structure of the real cities by means of spatial disaggregation.

## 3. Empirical evidence

The method of recursive subdivision of space and the model of hierarchical scaling can be easily applied to the cities in the real world by means of remote sensing data. Three megacities in Yangtze River delta, China, i.e., Shanghai, Nanjing, and Hangzhou, are taken as examples to show how to use the idea from spatial disaggregation to make scaling analysis (Figure 4). The remote sensing images used in this study came from National Aeronautics and Space Administration (NASA), including Landsat MSS, TM and ETM images of 1985, 1996 and 2005 of the three cities. These images were first rectified to a common UTM coordinate system based on appropriate topographic maps. The Supervised Classification method was employed to extract the built-up area with ERDAS IMAGINE software. And then, we modified the result artificially in ESRI ArcGIS to ensure better accuracy. After extracting the built-up patches, it is convenient to accomplish the procedure of spatial subdivision in ArcGIS. Using the *Intersect* function of ArcToolbox, we could extract the built-up area which fall into each divisional rectangular region, and then calculate areal values through the *Attribute Table*. Finally, we have we have nine datasets of urban land use for the three cities in three years.

The procedure of the urban spatial analysis is as follows. **Step 1: identify urban boundaries.** There are at least three approaches to designating metropolitan areas or demarcating urban agglomerations. The first is the city clustering algorithm (CCA) proposed by Rozenfeld *et al* (2008, 2011), the second is the fractal-based method presented by Tannier *et al* (2011), and the third is to derive what is called 'natural cities' by clustering street nodes/blocks (Jia and Jiang, 2011; Jiang and Jia, 2011). **Step 2: recursively subdivide urban space.** The algorithm of strict subdivision has been illustrated in Section 2. **Step 3: process data**. The datasets of areal size of parts in different rectangle frames can be picked up and processed stage by stage. **Step 4: perform scaling analysis**. Fitting the data to Zipf's law according to scale-free range, we can verify the theoretical model of intra-urban hierarchies.

The strict subdivision of urban space can be implemented by using the tools from ArcGIS. For example, for Shanghai in 2005, we have an urban area of 1418.4 km$^2$ after identifying the urban boundary and making a rectangle frame. Dividing the rectangle into two equal regions, we have two parts and the areal sizes are 814.229 km$^2$ and 604.176 km$^2$. Dividing the two smaller rectangles into four equal regions, we have four areal sizes: 470.4770 km$^2$, 380.4410 km$^2$, 343.7520 km$^2$, and 223.7340 km$^2$. The rest can be treated by analogy (Table 2). It is inevitable that the spatial subdivision gives rise to some errors so that the whole is not exactly equal to the sum of parts. The number of parts at each level is defined by the number law of urban hierarchy, equation (2). As a result, the average area can be perfectly fitted to the size law, equation (3), which can be readily verified by using the data displayed in Table 3. This suggests that the number and average area follow the hierarchical scaling law, and can be described with equation (4).

**Table 2 The results of the first three steps of spatial recursive subdivision of Shanghai, Nanjing, and Hangzhou in 1985, 1996, and 2005**

| City | Level | Area (sq.km.) | | |
|---|---|---|---|---|
| | ($m$) | 1985 | 1996 | 2005 |
| **Shanghai** | 1 | 520.0730 | 712.8160 | 1418.4000 |
| | 2 | 342.0090 | 469.0650 | 814.2290 |
| | | 178.0640 | 243.7510 | 604.1760 |
| | 3 | 175.4610 | 236.0090 | 470.4770 |
| | | 166.5490 | 233.0560 | 380.4410 |
| | | 112.1410 | 154.7550 | 343.7520 |
| | | 65.9223 | 88.9955 | 223.7340 |
| **Nanjing** | 1 | 172.1850 | 226.1640 | 339.4040 |
| | 2 | 104.3080 | 138.4600 | 203.5380 |
| | | 67.8765 | 87.7038 | 135.8650 |
| | 3 | 56.1852 | 76.4546 | 117.0610 |
| | | 48.1232 | 62.0059 | 86.4779 |
| | | 40.8913 | 51.3690 | 75.7399 |
| | | 26.9852 | 36.3349 | 60.1256 |
| **Hangzhou** | 1 | 79.9463 | 105.0030 | 185.4170 |
| | 2 | 47.8473 | 64.6180 | 122.8250 |
| | | 32.0990 | 40.3844 | 62.5924 |
| | 3 | 35.2971 | 41.2566 | 66.2761 |
| | | 16.0693 | 23.3615 | 56.5488 |
| | | 16.0297 | 21.0526 | 37.5807 |
| | | 12.5502 | 19.3319 | 25.0117 |

**Table 3 The number and average area of parts at the first ten levels of the urban hierarchies of Shanghai, Nanjing, and Hangzhou**

| $m$ | $f_m$ | $A_m$ (sq.km.) | | | | | | | | |
|---|---|---|---|---|---|---|---|---|---|---|
| | | Shanghai | | | Nanjing | | | Hangzhou | | |
| | | 1985 | 1996 | 2005 | 1985 | 1996 | 2005 | 1985 | 1996 | 2005 |
| 1 | 1 | 520.0730 | 712.8160 | 1418.4000 | 172.1850 | 226.1640 | 339.4040 | 79.9463 | 105.0030 | 185.4170 |
| 2 | 2 | 260.0365 | 356.4080 | 709.2025 | 86.0923 | 113.0819 | 169.7015 | 39.9732 | 52.5012 | 92.7087 |
| 3 | 4 | 130.0183 | 178.2039 | 354.6010 | 43.0462 | 56.5411 | 84.8511 | 19.9866 | 26.2507 | 46.3543 |
| 4 | 8 | 65.0091 | 89.1019 | 177.3006 | 21.5231 | 28.2705 | 42.4255 | 9.9933 | 13.1253 | 23.1772 |
| 5 | 16 | 32.5045 | 44.5510 | 88.6502 | 10.7615 | 14.1353 | 21.2127 | 4.9966 | 6.5627 | 11.5886 |
| 6 | 32 | 16.2522 | 22.2754 | 44.3249 | 5.3808 | 7.0676 | 10.6064 | 2.4983 | 3.2813 | 5.7943 |
| 7 | 64 | 8.1261 | 11.1377 | 22.1626 | 2.6904 | 3.5338 | 5.3032 | 1.2492 | 1.6407 | 2.8971 |

| 8 | 128 | 4.0631 | 5.5689 | 11.0813 | 1.3452 | 1.7669 | 2.6516 | 0.6246 | 0.8203 | 1.4486 |
| 9 | 256 | 2.0315 | 2.7844 | 5.5406 | 0.6726 | 0.8835 | 1.3258 | 0.3123 | 0.4102 | 0.7243 |
| 10 | 512 | 1.0158 | 1.3922 | 2.7703 | 0.3363 | 0.4417 | 0.6629 | 0.1561 | 0.2051 | 0.3621 |

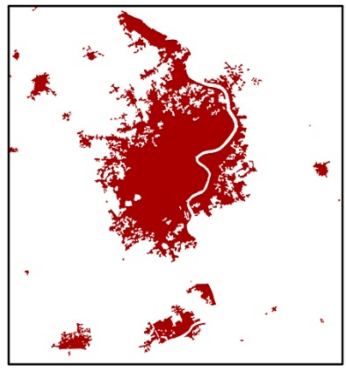 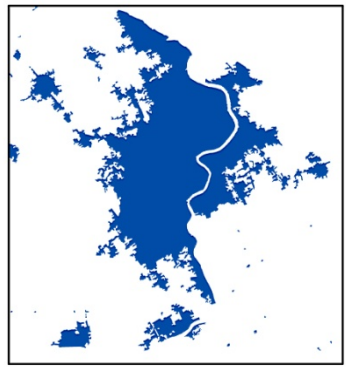 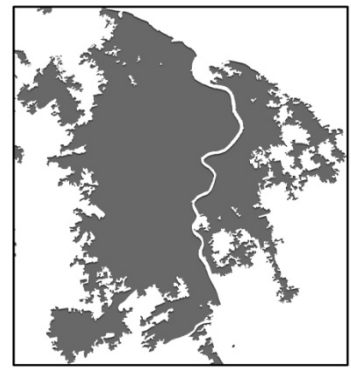

1985      1996      2005

a. Shanghai

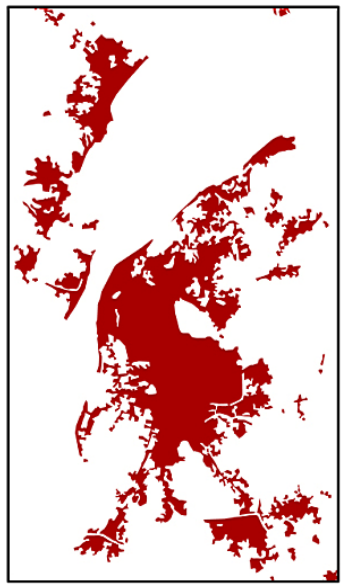 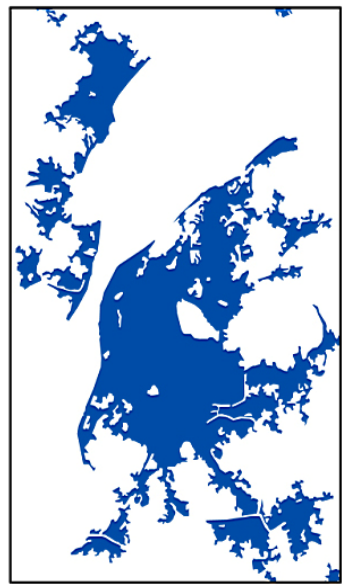 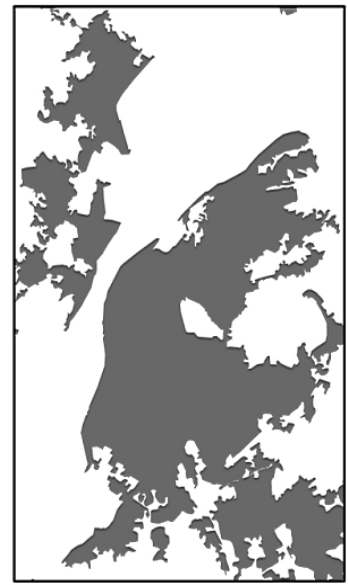

1985      1996      2005

b. Nanjing

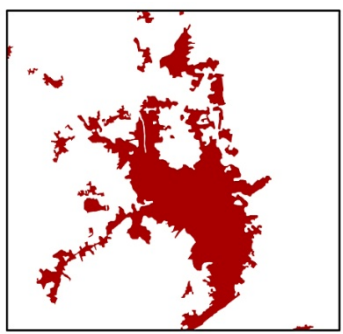 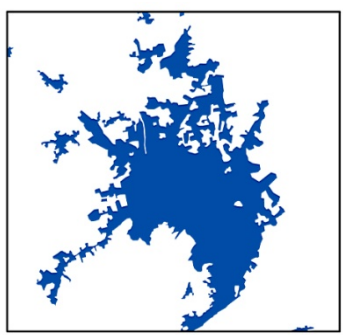 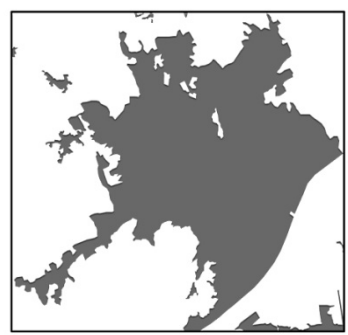

1985      1996      2005

c. Hangzhou

**Figure 4 The sketch maps of Shanghai, Nanjing, and Hangzhou in 1985, 1996, and 2005**

The key question of this work is whether or not the parts in the urban hierarchies comply with Zipf's law. After putting the numerical values in descending order by areal sizes, we can plot the rank and area on the double logarithmic graphs, with the axes being log (rank order) and log (areal size). The results show that there exists a typical scaling range in each plot. The first several parts diverge from the trend line of the scaling range in a sort, but the departure is not significant. However, the last parts depart from the trend line evidently and form a 'drooping tail'. As is often the case, the largest and smallest elements always make exceptions of the power-law distributions (Bak, 1996; Chen, 2011; Chen, 2012a; Zhou, 1995). If we remove the drooping tails at the ends, the data points can be fitted to equation (1), and the effect of statistical analyses is satisfying (Figure 5). For example, for Shanghai in 1985, the least squares computation of the first 2048 data points yields

$$\hat{A}_k = 1084.2280 k^{-1.0398},$$

where the hat '^' above $A_k$ denotes 'estimated value'. The goodness of fit is about $R^2=0.9929$, and the scaling exponent is $q=1.0398\pm0.0038$. For Nanjing in 1996, the least squares calculation of the first 2048 data points gives

$$\hat{A}_k = 373.0785 k^{-1.0032},$$

The goodness of fit is around $R^2=0.9932$, and the scaling exponent is $q=1.0032\pm0.0036$. For Hangzhou in 2005, the least squares computation of the first 4096 data points produces

$$\hat{A}_k = 302.2015 k^{-1.0089},$$

The goodness of fit is about $R^2=0.9756$, and the scaling exponent is $q=1.0089\pm0.0049$. The other datasets can be addressed with the similar method.

**Table 4 The parameters and the related statistics of Zipf's laws for Shanghai, Nanjing, and Hangzhou**

| Parameter | Shanghai | | | Nanjing | | | Hangzhou | | |
|---|---|---|---|---|---|---|---|---|---|
| | 1985 | 1996 | 2005 | 1985 | 1996 | 2005 | 1985 | 1996 | 2005 |
| $n$ | 2048 | 2048 | 4096 | 2048 | 2048 | 4096 | 2048 | 2048 | 4096 |
| $A_1$ | 1084.2280 | 1177.7026 | 2024.4880 | 270.5014 | 373.0785 | 548.6138 | 152.4563 | 164.3646 | 302.2015 |
| $q$ | 1.0398 | 1.0015 | 0.9891 | 0.9901 | 1.0032 | 1.0057 | 1.0246 | 0.9944 | 1.0089 |
| $R^2$ | 0.9929 | 0.9890 | 0.9853 | 0.9958 | 0.9932 | 0.9832 | 0.9888 | 0.9891 | 0.9756 |

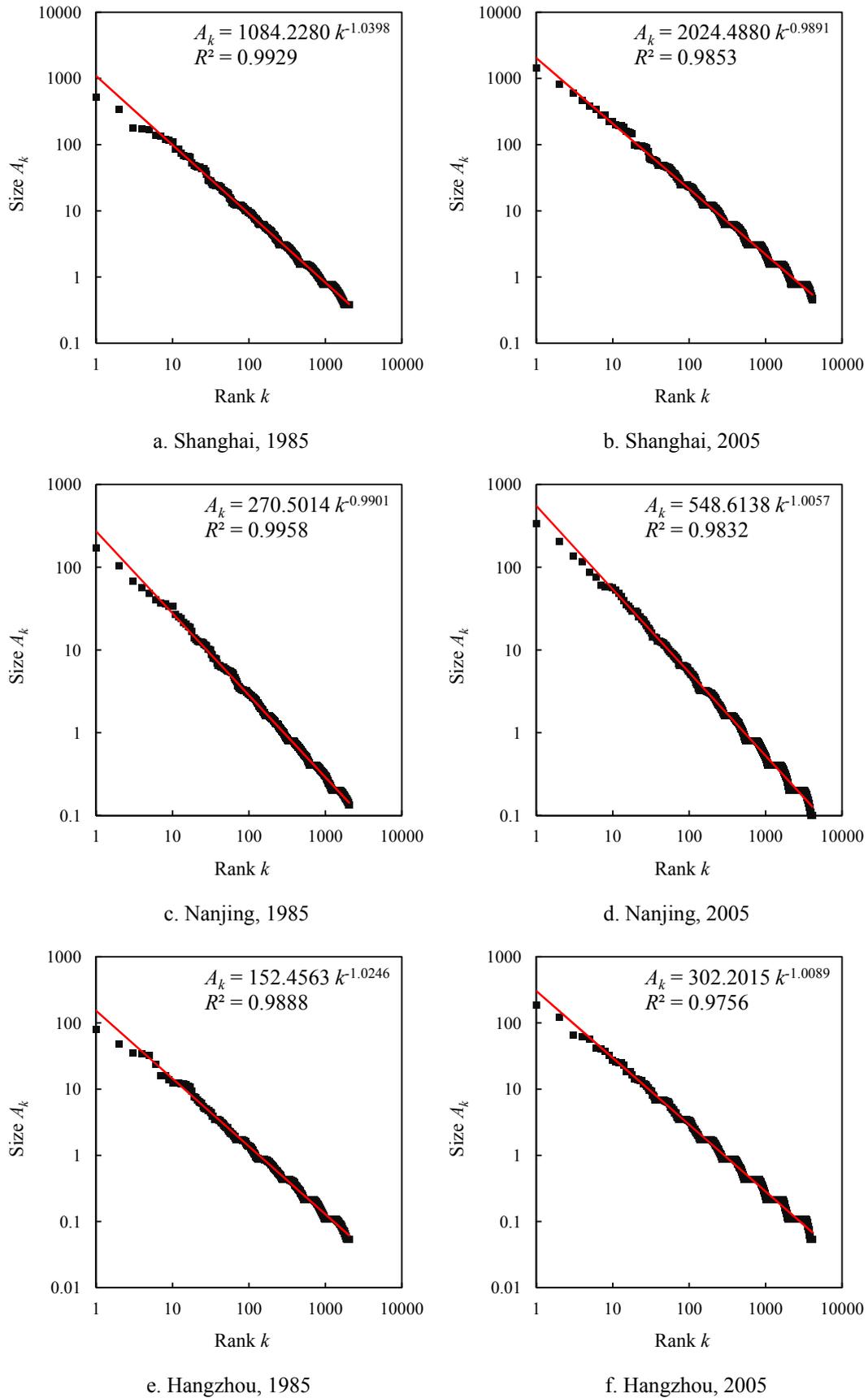

**Figure 5 The rank-size patterns of Shanghai, Nanjing, and Hangzhou in 1985 and 2005**

**Note**: The 'drooping tails' in the log-log plots have been removed. According to the scaling ranges of the rank-size distributions, the first 2048 data points are taken into account for 1985, and 4096 data points are taken for 2005.

For these three cities in three different years, the scaling exponent values are very close to $q$=1 (Figure 5; Table 4). In theory, the proportionality coefficient of the Zipf model, $A_1$, is expected to equal the total urban area. However, comparing Table 4 with Table 2 shows that there are significant differences between the observed values and the predicted values. For example, for Hangzhou in 2005, the urban area in the study region is about 185.417 km$^2$ (Table 2), but the proportionality constant of the model is $A_1$≈302.2015 (Table 4). This lends further support to the viewpoint that the largest element/part departs from the trend line of the power-law distribution (Bak, 1996; Chen, 2012d; Zhou, 1995). Despite this deficiency familiar to physical scientists, a conclusion can be drawn that the distribution of areal sizes in intra-urban hierarchies follows Zipf's law.

## 4. Questions and discussion

The above empirical analyses show that the recursive subdivision of space is an efficient approach to exploring spatial order of cities. The strict spatial subdivision is similar in form to but different in essence from the fractal-based box-counting method. The box-counting method has been employed to research growth patterns of cities (Benguigui *et al*, 2000; Feng and Chen, 2010; Shen, 2002). The strict subdivision is chiefly used to reveal rank-size scaling relation of space-filling processes of urban growth, not to estimate fractal dimension of urban form. The observational data of the three cities of China support the inference that intra-urban hierarchies follow Zipf's law. A circumstantial evidence for this inference is that the data resulting from Figure 2, the 'imaginary city', cannot be properly fitted to equation (1). This implies that it is real cities instead of random figures that satisfy the rank-size scaling relation.

The rank-size distribution is mathematical equivalent to hierarchical structure (Chen, 2012d). The urban hierarchy of internal structure of a city is different from the hierarchy of cities despite their common ground. This suggests that the rank-size distribution of cities differs to a certain extent from the Zipf distribution of strict subdivision results of intra-urban form (Table 5). For the intra-urban hierarchy within a city, the parts at the upper levels comprise the smaller parts at the lower levels. The relationships between the different levels are in fact the relationships between parts and the whole. For the interurban hierarchy consisting of different cities in a geographical region, the cities at the upper levels do not include the cities at the lower levels. The relationships between different levels are the relationships between different nodes in a network. In short, the interurban size distribution is to describe the hierarchical order of network of cities, while the intra-urban size distribution is to characterize the spatial organization of an individual city. This reminds us of fractals such as the Cantor set, which is a standard hierarchy. The fractal hierarchy is very similar to the intra-urban cascade structure instead of a hierarchy of cities. For a regular fractal, the relationships between different levels are also the relationships between parts and whole. The copies of a monofractal hierarchy of the Cantor set do not conform to Zipf's law, but a multifractal hierarchy of this set is consistent with the Zipf distribution (Chen, 2012b). In theory, the intra-urban hierarchy is based on dilation symmetry, while the interurban hierarchy is based on both dilation symmetry and translational symmetry. In spite of the differences between the intra-urban hierarchy and the interurban hierarchy, both of them can be modeled with the similar mathematical equations. The two kinds of hierarchies follow Zipf's law and can be described with the hierarchical scaling relations. This suggests that a city as a system is an epitome of a system of cities, and many theories and models on intra-urban structure can be generalized to interurban structure and *vice versa*.

Table 5 The similarities and differences between the interurban size distribution and the intra-urban size distribution

| Item | Interurban distribution | Intra-urban distribution |
| --- | --- | --- |
| **Description object** | Hierarchical order of spatial network | Spatial organization of urban form |
| **Element** | Different cities in a region | Different parts in a city |
| **Size distribution** | Zipf distribution | Zipf distribution |
| **Scaling exponent** | Close to 1 with large variation | Close to 1 with small variation |
| **Fractal property** | Monofractal or multifractal | Multifractals |
| **Cascade structure** | Mathematical construction | Physical structure |
| **Hierarchical relation** | The upper classes do not include the lower classes | The upper classes include the lower classes |
| **Symmetry** | Dilation and translational symmetry | Dilation symmetry |
| **Mathematical transform** | From rank-size distribution to hierarchical structure | From hierarchical structure to rank-size distribution |

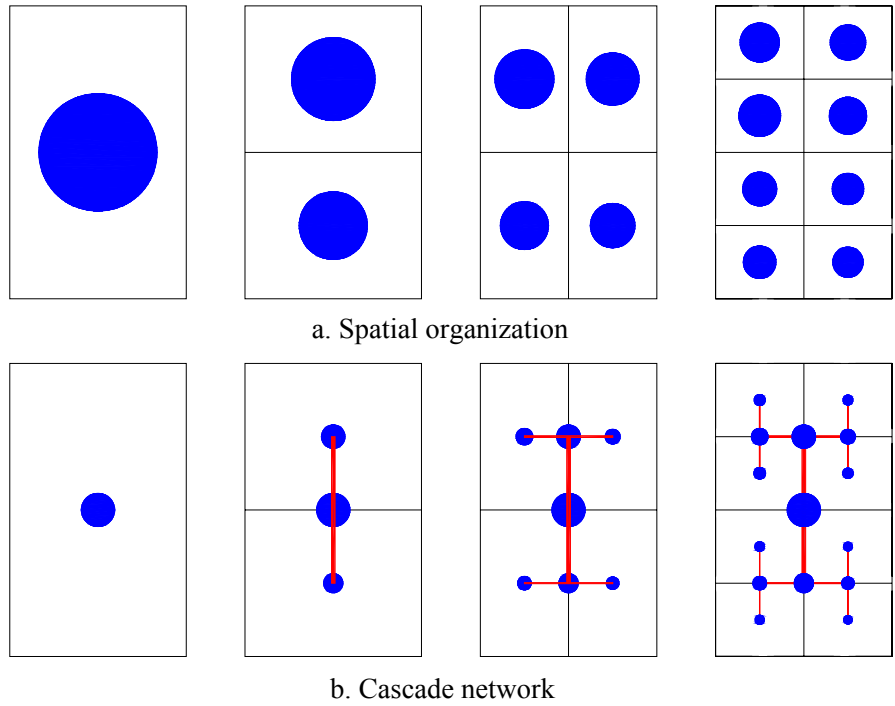

a. Spatial organization

b. Cascade network

**Figure 6 The spatial organization and network structure based on the rank-size distribution (the first four steps)**

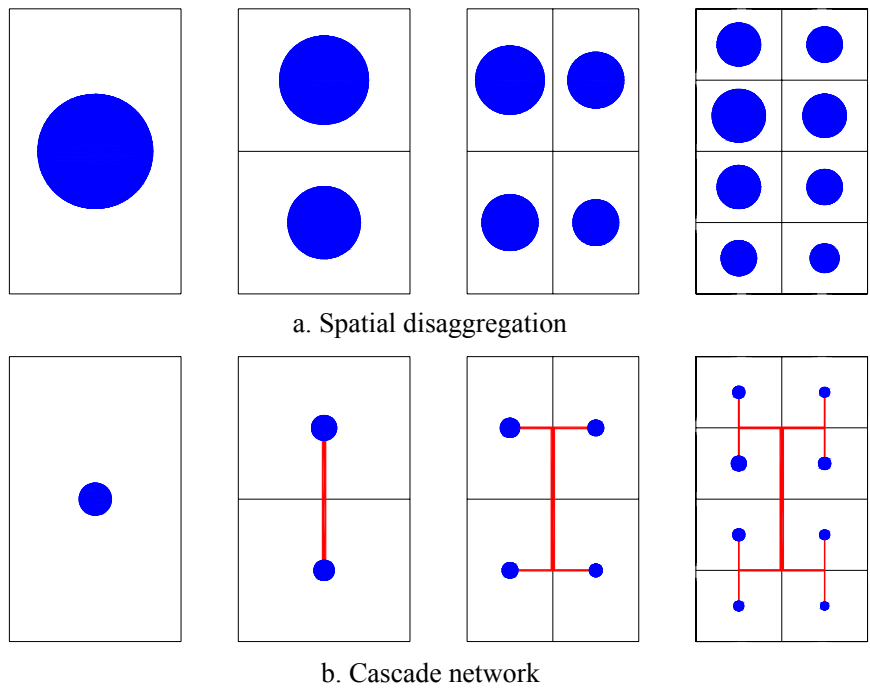

a. Spatial disaggregation

b. Cascade network

**Figure 7 The spatial disaggregation and network structure based on multifractal organization (the first four steps)**

Based on Zipf's law, the process of urban spatial disaggregation displayed in Figures 1 and 2

can be revised and abstracted as new models (Figure 6, Figure 7). These models provide a novel way of looking at urban space-filling. Urban growth is actually a process with scaling invariance. Moreover, the models are helpful for us to understand the spatial recursive subdivision, scale-free network, and, especially, self-similar hierarchy in the right perspective. The hierarchical scaling relations are here and there in human systems. Because of chance factors, we cannot find this regular structure from real cities and systems of cities. Through statistical analysis, the latent hierarchical scaling can be revealed through the Zipf distribution. A cards-shuffling model has been proposed to explain the relationships between the regular forms in the possible world and irregular patterns in the real world (Chen, 2012c). The physical order of urban systems can be revealed through mathematical laws.

However, if we investigate Table 2 carefully, we can find that the urban hierarchy is not always organized in keeping with the rank-size rule. The Zipf distribution falls into two groups: monofractal-based rank-size distribution and multifractal-based rank-size distribution (Chen, 2012b). If the hierarchical structure is determined by Zipf's law indicative of monofractal structure, the size of elements at the upper levels must be greater than those at the lower levels. The real urban hierarchies are not the case. For example, for Hangzhou in 2005, the areal size at the second level, 62.5924, is clearly less than a value at the third level, 66.2761. If we examine more levels, we can find more 'exceptional' arrangement in terms of the rank-size principle. For the real cities such as Shanghai and Nanjing, these 'abnormal' phenomena appear again and again, especially at the lower levels. If we want to fit the observational data to Zipf's law, we must ignore the real hierarchical structure and realign the rank of elements by areal size. The inconsistent relationship between Zipf's law and the hierarchical structure can be eliminated with the idea from multifractals (Chen, 2012b). That is, if we fit the data to the multifractal distribution rather than the monofractal distribution, the difficult problem will be readily resolved. The similarities and differences between the two distributions can be illustrated with Figures 6 and 7. In addition, a comparison between multifractal distribution and the Zipf distribution can be drawn through Table 6 (the first four steps). The concepts from multifractals have been applied to human systems (Appleby, 1996; Chen, 2012b; Haag, 1994). A multifractal model of size distribution will be put forward for intra-urban hierarchy in our future work.

**Table 6 Comparison between multifractal distribution with the rank-size distribution (the first four steps)**

| Example | $m$ | $f_m$ | Area | $A_m$ |
|---|---|---|---|---|
| Rank-size distribution ($q=1$) | 1 | 1 | 1.0000 | 1.0000 |
| | 2 | 2 | 0.5000, 0.3333 | 0.4167 |
| | 3 | 4 | 0.2500, 0.2000, 0.1667, 0.1429 | 0.1899 |
| | 4 | 8 | 0.1250, 0.1111, 0.1000, 0.0909, 0.0833, 0.0769, 0.0714, 0.0667 | 0.0907 |
| Multifractal distribution ($p=0.6$) | 1 | 1 | 1.000 | 1.000 |
| | 2 | 2 | 0.600, 0.400 | 0.500 |
| | 3 | 4 | 0.360, 0.240, 0.240, 0.160 | 0.250 |
| | 4 | 8 | 0.216, 0.144, 0.144, 0.096, 0.144, 0.096, 0.096, 0.064 | 0.125 |

**Note**: For the rank-size distribution, the scaling exponent $q=1$; For the two-scale multifractal distribution, the probability is $p=0.6$, and thus $1-p=0.4$.

## 5. Conclusions

The recursive subdivision of space used to be a pure theoretical concept in geographical analysis. In this paper, the idea from spatial disaggregation is turned into a practical method and applied to urban space. Thus the theoretical notion is generalized to empirical research. The urban internal structure is demonstrated to conform to the rank-size rule by spatial subdivision. The strict subdivision of urban space provides a new way of looking at city fractals, and the intra-urban rank-size distribution is helpful to understanding the conventional Zipf distribution of cities in a geographical region.

From this study, the main conclusions can be drawn as follows.

**First, the recursive subdivision of space is new approach to bringing to light the latent spatial order of urban form.** By this method, we find that urban growth is a scale-free process of space filling. Urban patterns always look fragmentary and irregular on the digital maps. There seems to be no order behind urban landscape. Fractal geometry is a powerful tool for revealing the spatial order under urban patterns, and fractal dimension is one of the important parameters to characterize urban structure. This paper shows that, besides box-counting method for fractal analysis, the spatial disaggregation is a useful tool to disclose the spatial regularity, and the scaling exponent of the rank-size distribution may be a significant measurement to reflect the

geographical feature of urban landscape.

**Second, the hierarchical scaling indicates more general structure than fractals and the rank-size distribution.** A fractal is a scaling phenomenon based on dilation symmetry, while the rank-size distribution is a scaling relation based on dilation symmetry and translational symmetry. Zipf's law indicative of the rank-size distribution is just the signature of self-similar hierarchy. There are similarities and differences between fractals, intra-urban strict subdivision, and hierarchy of cities, but all of these can be modeled with hierarchical scaling law. This implies that fractals, intra-urban hierarchy, and hierarchy of cities can be integrated into a new theoretical framework. The framework provides a possible approach to understanding complex systems such as cities and networks of cities.

**Third, the hierarchical structure of cities suggests possible multifractal distribution behind the recursive subdivision of urban space.** If the intra-urban hierarchy is of monofractal structure, the areal sizes at the upper levels must be greater than the sizes at the lower levels. However, for the results from the real cities, a few parts make an exception. Sometimes, the areal size of a part at the upper level is smaller than that at the lower level. This phenomenon cannot be modeled by the hierarchical scaling law based on the rank-size rule, but it can be modeled with the hierarchical scaling law based on multifractals. The multifractal structure of intra-urban hierarchy remains to be deeply researched in the future work.

## Acknowledgements

This research was sponsored by the National Natural Science Foundation of China (Grant No. 41171129).